\documentclass[11pt]{amsart}
\usepackage{amsmath, amsthm}
\numberwithin{equation}{section}

\begin{document}
\title{Analytic Partial Wave Expansion and Integral Representation of Bessel Beam}
\author{Amer Hodzic}
\address{Department of Physics, University of Rhode Island, Kingston, RI 02881, USA}

\begin{abstract}
This paper describes the partial wave expansion and integral representation of Bessel beams in free space and in the presence of dispersion. The expansion of the Bessel beam wavepacket with constant spectrum is obtained as well. Furthermore, the sum of a triple Legendre polynomial product of same order but different argument follows naturally from the partial wave expansion. The integration of all Bessel beams over all conical angles is shown to have a simple series representation, which confirms the equivalence between the results for both expansion and integral representation. 
\end{abstract}
\maketitle
\section{Introduction}

The Bessel beam is a mathematical construct that is a solution to wave equation. Bessel beams have attracted attention ever since Durnin \cite{Durnin} published his paper in 1987. The attention is due to the beam's intensity profile that has a sharp peak along the axis of propagation. The electromagnetic Bessel beam also travels with a phase velocity that is higher than speed of light. Numerous authors have used a superposition of Bessel beams in order to form a wavepacket (for example see \cite{Lu, Salo, Rached}). Their choice of using an exponential spectrum has allowed them to create a wavepacket that has been called an X-wave. The X-wave however is not a physically possible solution because it carries infinite energy. There is also interest in the superluminal property of X-waves. Two papers have been recently published that clearly show how  two schools of thought concerning the X-waves have opposing views. One side states that X-waves are physically realizable, even superluminally \cite{Rached 2}, whereas the other side states that superluminal effects of X-waves are simply phase related and that superluminal information transfer via X-waves is not possible \cite{Walker}. The author of this paper shares the latter view. Nevertheless, the mathematics behind Bessel beams, which are building blocks for wavepackets, is very rich as shown in this paper. The mathematical tools developed here may help create a proper wavepacket that has finite energy and is therefore physically realizable. Such a wavepacket may still be of interest because Bessel beam itself has reconstructive properties not recognized in other types of beams \cite{McGloin}. In this paper I will show and confirm two representations of Bessel beam, namely a partial wave expansion and an integral representation. The partial wave expansion of Bessel beam has been reported before, as will be explained later, but the proof presented here is new and based on a physical argument. The integral representation however is to my knowledge a new result. Furthermore I will show that a simple superposition of Bessel beams with a constant spectrum is suprisingly related to the infinite series of triple Legendre polynomial product. In the process I will show a Bessel beam identity that is simple and also new. I will also briefly describe the above two representations when dispersion is included into wave equation. 

\section{Mathematical Introduction}               
A zeroth order Bessel beam that propagates in the z direction can be described by
\begin{equation}
\label{Bessel Beam}
	\Phi(\rho,z,t)=\exp(ik_{z}z-\imath\omega t)J_{0}(k_{\rho}\rho)
\end{equation}
where $\rho$ is the transversal distance from the propagation axis $z$.

Solution (\ref{Bessel Beam}) satisfies the wave equation 
\begin{equation}
	\nabla^{2}\Phi=\frac{1}{c^{2}}\frac{\partial\Phi}{\partial t}
\end{equation}
where
 \begin{equation}k_{z}=\frac{\omega}{c}\cos(\theta)\end{equation}
 \begin{equation}k_{\rho}=\frac{\omega}{c}\sin(\theta),\end{equation}

\noindent which is a wave vector construct used by Salo et al. \cite{Salo}.

The angle $\theta$ is the axicon angle of the wave. The wave vector $k=(k_{z},k_{\rho},k_{\phi})$ is restricted to lie on a cone described by vertex angle $\theta$. The Bessel beam itself constitutes the sum of all wavevectors on this cone (i.e. sum over all $k_{\phi}$). For simplicity I shall let $c=1$.

I will first show that the partial wave expansion of a Bessel beam is
\begin{align}
\label{Partial Wave Expansion}
		&\exp[\imath\omega \cos(\theta)z-\imath\omega t] J_{0} [\omega \sin(\theta)\rho]		
		\nonumber\\
		&=\displaystyle\sum_{n=0}^\infty
		2\imath^{n}(n+\frac{1}{2})
		P_{n}[\cos(\theta)]
		P_{n}[\frac{z}{\sqrt{z^{2}+\rho^{2}}}]
		j_{n}(\omega\sqrt{z^{2}+\rho^{2}})
		\exp[-\imath\omega t],
\end{align}
where $P_{n}$ is Legendre polynomial of order $n$ and $j_{n}$ is a spherical Bessel function of same order. 

I will then show that the integral representation of Bessel beam is
\begin{align}
\label{Integral Representation}
	&\exp[\imath\omega \cos(\theta)z-\imath\omega t]
	J_{0}[\omega \sin(\theta)\rho]
	\nonumber\\
	&=\frac{1}{\pi}
	\displaystyle\int_{-\infty}^\infty
	j_{0}(R)
	\exp[\frac{\imath z \lambda}{\sqrt{z^{2}+\rho^{2}}}]
	\exp(-\imath\omega t)	
	d\lambda,
\end{align}
where $j_{0}$ is spherical Bessel function of order zero. $R$ is given by 
\begin{equation}
	R=\sqrt
	{\lambda^{2}+\omega^{2}(z^{2}+\rho^{2})-2\lambda\omega\sqrt{z^{2}+\rho^{2}}\cos(\theta)}.
\end{equation}

 The Bessel beam wavepacket is a superposition of Bessel beams:
\begin{equation}
	\Psi(\rho,z,t)=\displaystyle\int_{-\infty}^\infty
	S(\omega)\exp[\imath\omega \cos(\theta)z-\imath\omega t]
	J_{0}[\omega \sin(\theta)\rho]
	d\omega.
\end{equation}

\noindent If $S(\omega)=1$, I will show that 
\begin{align}
\label{Bessel beam wavepacket expansion}
		&\displaystyle\int_{-\infty}^\infty
		\exp[\imath\omega \cos(\theta)z-\imath\omega t]
		J_{0}[\omega \sin(\theta)\rho]		
		d\omega																		
		\nonumber\\
		&=\frac{1}{\sqrt{z^{2}+\rho^{2}}}
		\displaystyle\sum_{n=0}^\infty
		2\pi(n+\frac{1}{2})
		P_{n}[\cos(\theta)]
		P_{n}[\cos(\eta)]
		P_{n}[\cos(\gamma)].
\end{align}

\noindent where $\cos(\eta)\mbox{=\ensuremath{\frac{z}{\sqrt{z^{2}+\rho^{2}}}}}$ and $\cos(\gamma)=\frac{t}{\sqrt{z^{2}+\rho^{2}}}$.

The integral in (\ref{Bessel beam wavepacket expansion}) can be evaluated analytically, and from this I will show the result for the following triple Legendre polynomial sum
\begin{align}
\label{Triple Legendre Polynomial Sum Result}
		&\displaystyle\sum_{n=0}^\infty
		(2n+1)
		P_{n}[\cos(\theta)]
		P_{n}[\cos(\eta)]
		P_{n}[\cos(\gamma)]
		\nonumber\\
		&=\frac{2}
		{\sqrt{\sin^{2}(\eta) \sin^{2}(\theta)-
		[\cos(\eta)\cos(\theta)-\cos(\gamma)]^{2}}}
\end{align}
\noindent if $\sin(\eta)\sin(\theta)>\mid\cos(\eta)\cos(\theta)-\cos(\gamma)\mid$, otherwise the sum equals zero.

The result in (\ref{Triple Legendre Polynomial Sum Result}) is elegant, short and simply obtained. Similar result has been found by Baranov \cite{Baranov} by using different, more complicated and lenghty method. My result coincides with that of Baranov \cite{Baranov} and I have numerically verified it as well.

I will finally evaluate the following integral:  
\begin{equation}
	\int_{-1}^1
	J_{0}[\omega r\sin^{2}(\theta)]
	\exp[\imath\omega r\cos^{2}(\theta)]
	d[\cos(\theta)]
	=\sum_{n=0}^\infty
	2\imath^{n}j_{n}(\omega r).
\end{equation}

For this I will use both the partial wave expansion and the integral representation of Bessel beams. In both cases I get the same result and this verifies the equivalence between (\ref{Partial Wave Expansion}) and (\ref{Integral Representation}). The result itself constitutes a Bessel beam identity which to my knowledge is new.

Marston \cite{Marston} gives and numerically verifies the Bessel beam partial wave expansion in (\ref{Partial Wave Expansion}). Mitri \cite{Mitri} shows the Bessel beam partial wave expansion directly from the Gegenbauer result  published in 1899. I can however find the exact form of the expansion as in (\ref{Partial Wave Expansion}) in the work of Stratton \cite[page~413]{Stratton}.                                                                                                                                                                                                                                                                                                                                                                                                                                                                                               The proof of (\ref{Partial Wave Expansion}) in this paper is new, and is described from a physical perspective. The integral representation in (\ref{Integral Representation}) is to my knowledge new. 
   
\section{Bessel Beam Partial Wave Expansion}

Let me write a superposition of Bessel beams with varying conical angle $\theta$:
\begin{equation}
	\label{Conical Superposition}
	\int_{-1}^1
	B[\cos(\theta)]
	\exp[\imath\omega \cos(\theta)z-\imath\omega t ]
	J_{0}[\omega \sin(\theta)\rho]
	d[\cos(\theta)],
\end{equation}
where $B[cos(\theta)]$ is angular spectrum. The angular spectrum can then be expanded in terms of Legendre polynomials, namely 
\begin{equation}
	B[\cos(\theta)]
	=\sum_{n=0}^\infty
	a_{n}P_{n}[\cos(\theta)],
\end{equation}
where $P_{n}$ is Legendre polynomial of order $n$ and coefficients $a_{n}$ are constant.

If I choose the spectrum to be a delta function that eliminates all Bessel beams from the superposition except the one with conical angle $\theta_{0}$:
\begin{equation}
	B[\cos(\theta)]
	=\delta[\cos(\theta)-\cos(\theta_{0})],
\end{equation}

\noindent then one has to let $a_{n}=(n+\frac{1}{2})P_{n}[\cos(\theta_{0})]$, because the standard identity for Legendre polynomials is:
\begin{equation}
	\sum_{n=0}^\infty
	(n+\frac{1}{2})
	P_{n}[\cos(\theta_{0})]
	P_{n}[\cos(\theta)]
	=\delta[\cos(\theta)-\cos(\theta_{0})].
\end{equation}

I let $\alpha=\cos(\theta)$ and $\alpha_{0}=\cos(\theta_{0})$. At $t=0$ and with delta function angular spectrum, (\ref{Conical Superposition}) becomes:
\begin{equation}
\label{Construct Delta Function}
	\int_{-1}^1
	\sum_{n=0}^\infty
	(n+\frac{1}{2})
	P_{n}(\alpha_{0})
	P_{n}(\alpha)
	\exp(\imath\omega\alpha z)
	J_{0}(\omega\sqrt{1-\alpha^{2}}\rho)
	d\alpha.
\end{equation}

The expression in (\ref{Construct Delta Function}) is equal to the expression for Bessel beam at $t=0$:
\begin{align}
\label{Construct Integral}
		&\exp(\imath\omega\alpha_{0} z)
		J_{0}(\omega\sqrt{1-\alpha_{0}^{2}}\rho)
		\nonumber\\
		&=\displaystyle\int_{-1}^1
		\sum_{n=0}^\infty
		(n+\frac{1}{2})
		P_{n}(\alpha_{0})
		P_{n}(\alpha)
		\exp(\imath\omega\alpha z)
		J_{0}(\omega\sqrt{1-\alpha^{2}}\rho)
		d\alpha.
\end{align}

The right hand side of (\ref{Construct Integral}) is very usefull because the integral in  (\ref{Construct Integral})  can be evaluated, (see \cite{Stratton,Neves,Cregg,Koumandos,Dodonov}). I should mention that Neves et al. \cite{Neves} state that they do not know of any other report of this evaluation result. However Cregg and Svendlindh \cite{Cregg}, Koumandos \cite{Koumandos} and Dodonov \cite{Dodonov} clearly show that the result follows from publications of Gegenbauer at the end of 19th century. Nevertheless, Cregg and Svendlindh \cite{Cregg} state that the result is not generally well known. Stratton \cite[page~411]{Stratton} evaluates the integral in (\ref{Integral of Stratton}) in the exact form as well as authors in \cite{Neves,Cregg,Koumandos,Dodonov}:
\begin{align}
		&\label{Integral of Stratton}
		\int_{-1}^1
		P_{n}(\alpha)
		\exp(\imath\omega\alpha z)
		J_{0}(\omega\sqrt{1-\alpha^{2}}\rho)
		d\alpha
		\nonumber\\
		&=2\imath^{n}
		P_{n}[\frac{z}{\sqrt{z^{2}+\rho^{2}}}]
		j_{n}(\omega\sqrt{z^{2}+\rho^{2}}).
\end{align}
Equation (\ref{Construct Integral}) therefore becomes the partial wave expansion of Bessel beam written below for any time $t$:
\begin{align}
		&\exp(\imath\omega\alpha_{0}z-\imath\omega t)
		J_{0}(\omega\sqrt{1-\alpha_{0}^{2}}\rho)
		\nonumber\\
		&=\displaystyle\sum_{n=0}^\infty
		2\imath^{n}(n+\frac{1}{2})
		P_{n}(\alpha_{0})
		P_{n}[\mbox{\ensuremath{\frac{z}{\sqrt{z^{2}+\rho^{2}}}}}]
		j_{n}(\omega\sqrt{z^{2}+\rho^{2}})
		\exp(-\imath\omega t).
\end{align}

\section{Bessel Beam Integral Representation}

I can simplify the Bessel beam partial wave expansion in (\ref{Partial Wave Expansion}) at $t=0$ if I let $\beta\mbox{=\ensuremath{\frac{z}{\sqrt{z^{2}+\rho^{2}}}}}$ and $\mu=\omega\sqrt{z^{2}+\rho^{2}}$, to obtain
\begin{equation}
	\label{Partial Wave Expansion at time zero 1}
	\exp(\imath\omega\alpha_{0}z)
	J_{0}(\omega\sqrt{1-\alpha_{0}^{2}}\rho)=
	\displaystyle\sum_{n=0}^\infty
  	2\imath^{n}(n+\frac{1}{2})
	P_{n}(\alpha_{0})
	P_{n}(\beta)
	j_{n}(\mu).
\end{equation}

The Fourier transform relationship between Legendre polynomial $P_{n}(\beta)$ and the spherical Bessel function  $j_{n}(\lambda)$ is given by Mehrem \cite{Mehrem}:
\begin{equation}
	P_{n}(\beta)=
	\frac{(-\imath)^n}{\pi}
	\displaystyle\int_{-\infty}^\infty
	j_{n}(\lambda)
	\exp(\imath\beta\lambda)
	d\lambda.
\end{equation}

I can therefore further rewrite equation (\ref{Partial Wave Expansion at time zero 1}) as
\begin{align}
\label{Partial Wave Expansion at time zero 2}
	&\exp(\imath\omega\alpha_{0}z)
	J_{0}(\omega\sqrt{1-\alpha_{0}^{2}}\rho)
	\nonumber\\
	&=\frac{2}{\pi}\displaystyle\int_{-\infty}^\infty
	\sum_{n=0}^\infty
	(n+\frac{1}{2})
	P_{n}(\alpha_{0})
	j_{n}(\lambda)
	j_{n}(\mu) 
	\exp(\imath\beta\lambda)
	d\lambda.
\end{align}

The spherical Bessel function is defined as $j_{n}(\lambda)=\sqrt{\frac{\pi}{2\lambda}}J_{n+\frac{1}{2}}(\lambda)$,  where $J_{n+\frac{1}{2}}(\lambda)$ is the Bessel function of half integer order. With this, (\ref{Partial Wave Expansion at time zero 2}) becomes
\begin{align}
\label{Construct Sum}
		&\exp(\imath\omega\alpha_{0}z)
		J_{0}(\omega\sqrt{1-\alpha_{0}^{2}}\rho)
		\nonumber\\
		&=\displaystyle\int_{-\infty}^\infty
		\left\{\sum_{n=0}^\infty
		(n+\frac{1}{2})
		P_{n}(\alpha_{0})
		\frac{J_{n+\frac{1}{2}}(\lambda)}{\sqrt{\lambda}}
		\frac{J_{n+\frac{1}{2}}(\mu)}{\sqrt{\mu}}\right\} 
		\exp(\imath\beta\lambda)d\lambda.
\end{align}

The equation (\ref{Construct Sum}) is very useful because the sum within is evaluated by Hochstadt \cite[page~221]{Hochstadt}),
\begin{equation}
	\sum_{n=0}^\infty
	(n+\frac{1}{2})
	P_{n}(\alpha_{0})
	\frac{J_{n+\frac{1}{2}}(\lambda)}{\sqrt{\lambda}}
	\frac{J_{n+\frac{1}{2}}(\mu)}{\sqrt{\mu}}
	=\frac{1}{\pi}j_{0}(R),
\end{equation}

\noindent where $R$ is given by Hochstadt \cite[page~219]{Hochstadt} 
\begin{align}
	&R=
	\sqrt{\lambda^{2}+\mu^{2}-2\lambda\mu\alpha_{0}}
	\nonumber\\
	&=\sqrt{\lambda^{2}+\omega^{2}(z^{2}+\rho^{2})
	-2\lambda\omega\sqrt{z^{2}
	+\rho^{2}}\cos(\theta_{0})}.
\end{align}

I have therefore reduced equation (\ref{Construct Sum}) to the integral representation of Bessel beams written below for any time $t$.
\begin{align}
	&\exp(\imath\omega\alpha_{0} z-\imath\omega t)
	J_{0}(\omega\sqrt{1-\alpha_{0}^{2}}\rho)
	\nonumber\\
	&=\frac{1}{\pi}
	\displaystyle\int_{-\infty}^\infty
	j_{0}(R)
	\exp[\frac{\imath z \lambda}{\sqrt{z^{2}+\rho^{2}}}]
	\exp(-\imath\omega t)
	d\lambda.
\end{align}

\section{Expansion of Bessel Beam Wavepacket with Constant Spectrum}

Bessel beam wavepacket with constant spectrum can be expanded into an infinite series of product of three Legendre polynomials.  For this purpose let us integrate the partial wave expansion from (\ref{Partial Wave Expansion}) over variable $\omega$ in whole frequency domain:
\begin{align}
\label{Frequency Domain}
	&\displaystyle\int_{-\infty}^\infty
	\exp[\imath\omega \cos(\theta)z-\imath\omega t]
	J_{0}[\omega \sin(\theta)\rho]
	d\omega		
	\\
	&=\displaystyle\sum_{n=0}^\infty
	2\imath^{n}(n+\frac{1}{2})
	P_{n}[\cos(\theta)]
	P_{n}[\mbox{\ensuremath{\frac{z}{\sqrt{z^{2}+\rho^{2}}}}}]
	\displaystyle\int_{-\infty}^\infty
	j_{n}(\omega\sqrt{z^{2}+\rho^{2}})
	\exp(-\imath\omega t)\nonumber
	d\omega.
\end{align}

I also make the following variable substitutions
\begin{equation}
	\cos(\eta)\mbox{=\ensuremath{\frac{z}{\sqrt{z^{2}+\rho^{2}}}}}
\end{equation}
\begin{equation}
	\omega=\frac{\mu}{\sqrt{z^{2}+\rho^{2}}}.
\end{equation}

The right hand side of (\ref{Frequency Domain}) then becomes
\begin{equation}
\label{Partial Wave Fourier Transform}
	\frac{1}{\sqrt{z^{2}+\rho^{2}}}
	\sum_{n=0}^\infty
	2\imath^{n}(n+\frac{1}{2})
	P_{n}[\cos(\theta)]
	P_{n}[\cos(\eta)]
	\displaystyle\int_{-\infty}^\infty
	j_{n}(\mu)
	\exp[\frac{-\imath\mu t}{\sqrt{z^{2}+\rho^{2}}}]
	d\mu.
\end{equation}

I recognize the integral inside (\ref{Partial Wave Fourier Transform}) as Fourier transform of Legendre polynomial namely
\begin{equation}
P_{n}[\frac{t}{\sqrt{z^{2}+\rho^{2}}}]
	=\frac{1}{\pi(-\imath)^n}
	\displaystyle\int_{-\infty}^\infty
	j_{n}(\mu)
	\exp[\frac{-\imath\mu t}{\sqrt{z^{2}+\rho^{2}}}]
	d\mu.
\end{equation}

I let $cos(\gamma)=\frac{t}{\sqrt{z^{2}+\rho^{2}}}$ and therefore (\ref{Frequency Domain}) becomes the desired expansion
\begin{align}
\label{X Wave Expansion}
		&\displaystyle\int_{-\infty}^\infty
		\exp[\imath\omega \cos(\theta)z-\imath\omega t]
		J_{0}[\omega \sin(\theta)\rho]
		d\omega		
		\nonumber\\	
		&=\frac{1}{\sqrt{z^{2}+\rho^{2}}}
		\displaystyle\sum_{n=0}^\infty
		2\pi(n+\frac{1}{2})
		P_{n}[\cos(\theta)]
		P_{n}[\cos(\eta)]
		P_{n}[\cos(\gamma)].
\end{align}

The left hand side of (\ref{X Wave Expansion}) is a superposition of all Bessel beams with frequencies $\omega$ from $-\infty$ to $\infty$, and whose frequency spectrum is $f(\omega)=1$. The integral on the left hand side of (\ref{X Wave Expansion}) can be evaluated analytically as well and the result is:
\begin{align}
\label{X Wave Integral}
	&\displaystyle\int_{-\infty}^\infty
	\exp[\imath\omega \cos(\theta)z-\imath\omega t]
	J_{0}[\omega \sin(\theta)\rho]
	d\omega
	\nonumber\\
	&=\frac{2}
	{\sqrt{\sin^{2}(\theta)\rho^{2}-[t-\cos(\theta)z]^{2}}}.
\end{align}	

The integral result in equation (\ref{X Wave Integral}) is valid if $\mid t-\cos(\theta)z \mid<\sin(\theta)\rho$, otherwise the integral is equal to zero. This means that the expansion in (\ref{X Wave Expansion}) vanishes if $\sin(\eta)\sin(\theta)<\mid \cos(\eta)\cos(\theta)-\cos(\gamma)\mid$. I therefore have evaluated the sum of the triple Legendre polynomial product as well: 
\begin{align}
	&\sum_{n=0}^\infty
	2\pi(n+\frac{1}{2})
	P_{n}[\cos(\theta)]
	P_{n}[\cos(\eta)]
	P_{n}[\cos(\gamma)]
	\nonumber\\
	&=\frac{2\sqrt{z^{2}+\rho^{2}}}{\sqrt{\sin^{2}(\theta)\rho^{2}-[t-\cos(\theta)z]^{2}}}.
\end{align}

If I use the definitions for $\cos(\gamma)$ and $\cos(\eta)$, I finally get
\begin{align}
\label{Triple Legendre Polynomial Sum Proof}
		&\displaystyle\sum_{n=0}^\infty
		2\pi(n+\frac{1}{2})
		P_{n}[\cos(\theta)]
		P_{n}[\cos(\eta)]
		P_{n}[\cos(\gamma)]
		\nonumber\\
		&=\frac{2}{\sqrt{\sin^{2}(\eta)\sin^{2}(\theta)
		-[\cos(\eta)\cos(\theta)-\cos(\gamma)]^{2}}}.
\end{align}
The result of (\ref{Triple Legendre Polynomial Sum Proof}) is valid if $\sin(\eta)\sin(\theta)>\mid \cos(\eta)\cos(\theta)-\cos(\gamma)\mid$, otherwise the sum is equal to zero. The expansion of Bessel beam wavepacket with constant spectrum is a new result as well.

\section{Bessel Beam Identity}
In order to show that the two representations in (\ref{Partial Wave Expansion}) and (\ref{Integral Representation}) of Bessel beam are equivalent, I shall consider the following case: I will evaluate both representations along points of the cone only at $t=0$ and I will integrate the results over all possible cones. Let me start with expansion first. At a point on the cone I have $\cos(\eta)=\cos(\theta)$. I let $r=\sqrt{z^{2}+\rho^{2}}$ and I write:
\begin{align} 
		&\displaystyle\int_{-1}^1\sum_{n=0}^\infty
		J_{0}[\omega r\sin^{2}(\theta)]
		\exp[\imath\omega r\cos^{2}(\theta)]
		d[\cos(\theta)]
		\nonumber\\
		&=\displaystyle\int_{-1}^1
		\sum_{n=0}^\infty
		2\imath^{n}(n+\frac{1}{2})
		j_{n}(\omega r)
		P_{n}[\cos(\theta)]
		P_{n}[\cos(\theta)]
		d[\cos(\theta)].
\end{align}

If I exploit the orthogonality of Legendre polynomials I get
\begin{align}
\label{Conical Integral from Partial Wave Expansion}
		&\displaystyle\sum_{n=0}^\infty
		2\imath^{n}(n+\frac{1}{2})j_{n}(\omega r)
		\int_{-1}^1
		P_{n}[\cos(\theta)]
		P_{n}[\cos(\theta)]
		d[\cos(\theta)]
		\nonumber\\
		&=\displaystyle\sum_{n=0}^\infty
		2\imath^{n}(n+\frac{1}{2})
		j_{n}(\omega r)
		\frac{1}{(n+\frac{1}{2})}
		\nonumber\\
		&=\displaystyle\sum_{n=0}^\infty
		2\imath^{n}j_{n}(\omega r).
\end{align}

Similarly, let me proceed wih integral representation:
\begin{align}
\label{Integral Representation on Cone}
		&\displaystyle\int_{-1}^1
		J_{0}[\omega r\sin^{2}(\theta)]
		\exp[\imath\omega r\cos^{2}(\theta)]
		d[\cos(\theta)]
		\nonumber\\
		&=\displaystyle\int_{-1}^1
		\frac{1}{\pi}
		\displaystyle\int_{-\infty}^\infty
		j_{0}(R)
		\exp[\frac{\imath z \lambda}{\sqrt{z^{2}+\rho^{2}}}]
		d\lambda
	 	d[\cos(\theta)]
	 	\nonumber\\
		&=\displaystyle\int_{-1}^1
		\frac{1}{\pi}
		\displaystyle\int_{-\infty}^\infty\
		j_{0}(R)
		\exp[\imath\lambda \cos(\gamma)]
		d\lambda 
		d[\cos(\theta)].
\end{align}

The well known plane wave partial wave expansion namely \cite{Marion}
\begin{equation}
	\exp[\imath\lambda \cos(\gamma)]
	=\sum_{n=0}^\infty
	(n+\frac{1}{2})\imath^{n}
	j_{n}(\lambda)
	P_{n}[\cos(\gamma)]
\end{equation}
is used to rewrite the last line of (\ref{Integral Representation on Cone}) as:
\begin{align}
	\label{Gradshteyn Integral}
	&\displaystyle\int_{-1}^1
	\frac{1}{\pi}
	\displaystyle\int_{-\infty}^\infty
	j_{0}(R)\left\{ \sum_{n=0}^\infty
	(n+\frac{1}{2})\imath^{n}
	j_{n}(\lambda)
	P_{n}[\cos(\gamma)]\right\}
	d\lambda 
	d[\cos(\theta)]
	\nonumber\\
	&=\frac{1}{\pi}
	\displaystyle\sum_{n=0}^\infty
	(n+\frac{1}{2})\imath^{n}
	\displaystyle\int_{-\infty}^\infty
	j_{n}(\lambda)
	\left\{\displaystyle\int_{-\infty}^\infty
	j_{0}(R)P_{n}[\cos(\gamma)]
	d[\cos(\theta)]\right\} 
	d\lambda.
\end{align}

The inner integral of the right hand side of equation  (\ref{Gradshteyn Integral}) can be evaluated \cite{Gradshteyn} and therefore (\ref{Gradshteyn Integral}) reduces to:
\begin{equation}
	\frac{1}{\pi}
	\sum_{n=0}^\infty
	(n+\frac{1}{2})\imath^{n}
	\displaystyle\int_{-\infty}^\infty
	j_{n}(\lambda)
	\left\{ \frac{\pi}{\sqrt{\lambda\omega r}}
	J_{n+\frac{1}{2}}(\lambda)
	J_{n+\frac{1}{2}}(\omega r)\right\} 
	d\lambda.
\end{equation}

Further simplification leads to
\begin{align}
\label{Simplification}
		&\displaystyle\sum_{n=0}^\infty
		(n+\frac{1}{2})\imath^{n}
		j_{n}(\omega r)
		\displaystyle\int_{-\infty}^\infty
		\frac{[J_{n+\frac{1}{2}}(\lambda)] ^{2}}{\lambda}
		d\lambda
		\nonumber\\
		&=\displaystyle\sum_{n=0}^\infty
		(n+\frac{1}{2})\imath^{n}
		j_{n}(\omega r)\frac{2}{\pi}
		\displaystyle\int_{-\infty}^\infty
		j_{n}(\lambda)^{2}
		d\lambda.
\end{align}
The last integral in (\ref{Simplification}) can be evaluated as well:
\begin{equation}
	\displaystyle\int_{-\infty}^\infty
	j_{n}(\lambda)^{2}d\lambda
	=\frac{\pi}{2n+1}.
\end{equation}

My final result therefore is
\begin{equation}
\label{Bessel Beam Identity Result}
	\int_{-1}^1
	J_{0}[\omega r\sin^{2}(\theta)]
	\exp[\imath\omega r\cos^{2}(\theta)]
	d[\cos(\theta)]
	=\displaystyle\sum_{n=0}^\infty
	2\imath^{n}j_{n}(\omega r)
\end{equation}
which is the same result as in (\ref{Conical Integral from Partial Wave Expansion}). The equivalence between the partial wave expansion and integral representation of Bessel beam is therefore verified. The result in (\ref{Bessel Beam Identity Result}) constitutes a new Bessel beam identity.

\section{Partial Wave Expansion and Integral Representation with Dispersion}

The importance of the partial wave expansion of Bessel beam becomes more obvious when dispersion is allowed into the wave equation. Dispersion arises when the index of refraction depends upon frequency. In this case the Bessel beam partial wave expansion becomes
 \begin{align}
\label{Partial Wave Expansion with Dispersion}
		&\exp[\imath n(\omega)\omega \cos(\theta)z-\imath\omega t] J_{0} [n(\omega)\omega \sin(\theta)\rho]		
		\\
		&=\displaystyle\sum_{n=0}^\infty
		2\imath^{n}(n+\frac{1}{2})
		P_{n}[\cos(\theta)]
		P_{n}[\frac{z}{\sqrt{z^{2}+\rho^{2}}}]
		j_{n}[n(\omega)\omega\sqrt{z^{2}+\rho^{2}}]
		\exp[-\imath\omega t]\nonumber,
\end{align}
where $n(\omega)$ is frequency dependent index of refraction.

The index of refraction in the left hand side of (\ref{Partial Wave Expansion with Dispersion}) appears inside the Bessel function as well as inside the exponential, whereas $n(\omega)$ appears only inside the spherical Bessel function in the right hand side of (\ref{Partial Wave Expansion with Dispersion}). If a certain form of $n(\omega)$ allows the summation in (\ref{Partial Wave Expansion with Dispersion}) to be terminated at a certain integer value, then this may allow for better approximations. In the case of integral representation $n(\omega)$ is present inside the variable $R$ only:
\begin{equation}
R=\sqrt{\lambda^{2}+n^{2}(\omega)\omega^{2}(z^{2}+\rho^{2})
	-2\lambda n(\omega)\omega\sqrt{z^{2}
	+\rho^{2}}\cos(\theta)}.
\end{equation}

\section{Conclusion}
In this paper, I gave a new proof for partial wave expansion of a Bessel beam, from a physical perspective. The partial wave expansion has been shown by Stratton \cite[page~413]{Stratton} in a more general form. To my knowledge the integral representation of Bessel beam however is not found in literature. Furthermore the equivalency between the Bessel beam wavepacket with constant spectrum and the triple Legendre polynomial series as shown in section 5 is a new result as well. The similar sum has been found by Baranov \cite{Baranov} by using different more complicated method and our result coincides. The final result of  section 6 further shows the equivalence between the partial wave expansion and integral representation of Bessel beams, which further confirms the integral representation. I have further developed the mathematical structure of Bessel beams with the hope that the experimental results in Bessel beam optics can be described more accurately and the controversy of superluminal propagation resolved.

\end{document}